\documentclass[12pt]{article}
\usepackage{amsmath,amssymb,epsfig}

\usepackage{color}
\input{colordvi.tex}
\def\unit{{\relax{\rm 1\kern-.26em I}}}

\makeatletter

\renewcommand\section{\@startsection {section}{1}{\z@}%
                                   {-3.5ex \@plus -1ex \@minus -.2ex}%
                                   {2.3ex \@plus.2ex}%
                                   {\normalfont\large\bfseries}}

\renewcommand\subsection{\@startsection{subsection}{2}{\z@}%
                                     {-3.25ex\@plus -1ex \@minus -.2ex}%
                                     {1.5ex \@plus .2ex}%
                                     {\normalfont\normalsize\bfseries}}

\makeatother

\begin{document}

\baselineskip=18pt  
\numberwithin{equation}{section}  
\allowdisplaybreaks  



%
%


\thispagestyle{empty}

\vspace*{-2cm}
\begin{flushright}
\end{flushright}

\begin{flushright}
KYUSHU-HET-148
\end{flushright}

\begin{center}

\vspace{1.4cm}

\vspace{1cm}
{\bf\Large Decay of False Vacuum via }
\vspace*{0.3cm}

{\bf\Large Fuzzy Monopole in String Theory}
\vspace*{0.2cm}

\vspace{1.3cm}

{\bf
Aya Kasai$^{1}$ and  Yutaka Ookouchi$^{2,1}$} \\
\vspace*{0.5cm}

${ }^{1}${\it Department of Physics, Kyushu University, Fukuoka 810-8581, Japan  }\\
${ }^{2}${\it Faculty of Arts and Science, Kyushu University, Fukuoka 819-0395, Japan  }\\

\vspace*{0.5cm}

\end{center}

\vspace{1cm} \centerline{\bf Abstract} \vspace*{0.5cm}

We investigate dielectric branes in false vacua in Type IIB string theory. The dielectric branes are supported against collapsing by lower energy vacua inside spherical or tube-like branes. We claim that such branes can be seeds for semi-classical (or quantum mechanical) decay of the false vacua, which makes the life-time of the false vacua shorter. Also, we discuss a topology change of a bubble corresponding to the fuzzy monopole triggered by dissolving fundamental strings. 

\newpage
\setcounter{page}{1} 



\section{Introduction}

Existence of consistent flux compactifications in string theories may suggest us that there is a huge string landscape \cite{landscape1,landscape2}, and that there exists a large number of metastable vacua. If this is true, it would be important to study how transitions between them occur, and how long such metastable vacua survive\footnote{Remarkable studies on this subject have been done in various context (for example, see \cite{work1,work2,work3,Addedwork} and references therein). We would like to shed new light on this subject from a slightly different point of view. }. A key ingredient to answer this question may be a D-brane. In various scenarios of string model building, the standard model is realized as a low-energy effective theory of D-branes. So, it would be plausible to assume that lower dimensional D-branes also exist because the tachyon condensation of the original branes and anti-branes simply generates such lower dimensional D-branes. The lower dimensional D-branes can be interpreted as solitons in the effective theory. In this paper, we study some aspects of the solitonic branes in metastable states. Especially, we claim that such solitons induce the domain wall connecting the false vacuum and the true vacuum, and create a dielectric brane which can be interpreted as a bound state of the domain wall and the soliton. The dielectric branes are blown-up even when there is no background flux nor angular momentum. The stability of the branes is offered by the true vacuum inside spherical or tube-like branes. Our first aim of this paper is to point out new way to form a dielectric brane.

The other interesting observation of dielectric branes in metastable vacua can be seen in the decay process of the vacua. In a wide range of parameter space, such dielectric branes become unstable even when the false vacuum itself is stable. In this case, the radius of the dielectric brane expands rapidly without bound, and finally the whole four dimensional space-time is filled by a lower energy vacuum. This is a semi-classical decay process of the metastable vacua since this does not require the tunneling process. Decay of a metastable vacuum via dielectric branes was initially studied in \cite{KPV} where anti-D3 branes were blown-up to a NS5 brane and decayed into a supersymmetric vacuum\footnote{See \cite{work4} for recent interesting studies on instability of metastable states in this geometry.}. A crucial difference between discussions in \cite{KPV} and our studies is inhomogeneous vacuum decay. In \cite{KPV}, vacuum decayed by creating $O(4)$ symmetric bubble in the homogeneous space-time $ \mathbb{R}^{1,3}$. In our case, there are impurities (such as a monopole or a cosmic string) in $ \mathbb{R}^{1,3}$, and they enforce to create less-symmetric bubbles inhomogeneously. Existence of solitons reduces the energy cost to blow-up a domain wall connecting the metastable vacuum and true vacuum, and reduce the total energy by constructing a bound state of solitons and the domain wall. This idea was originally claimed in the context of field theory \cite{Pole,Hosotani,Yajnik}, then applied to phenomenological studies \cite{Kumar,Rstring,Lee}. The second aim of this paper is to go a step further toward applications of the idea to string theories. 

In this paper, to get insight into transitions between vacua in string landscape, we focus on simplified version of the landscape by taking the brane limit where $g_s$ is very small but $l_s$ is finite. As an example, we consider generalized conifold and study metastable vacuum engineered by wrapping D5 and anti-D5 branes \cite{Vafa}. Interestingly, in the metastable states, there is a monopole-like D-brane configuration which offers a good example to illustrate our idea\footnote{Similar kind of stringy monopole in the geometry \cite{KS} is pointed out first by \cite{StringyMonopole}.}.

The plan of this paper is the following. In section 2, we briefly review metastable states in Type IIB string theory compactified on the generalized conifold in the brane limit. In section 3, we discuss a new type of dielectric brane which is a bound state of a monopole and a domain wall connecting two states. In section 4, we calculate decay rate of the metastable bound state by means of the bounce solution of Euclidean Dirac-Born-Infeld (DBI) action. In section 5, we discuss topology change of the bound state triggered by colliding fundamental strings. In section 6, we point out that the warping effect drastically enhances instability of the bound state. Section 7 is devoted to the conclusions and discussions.

\section{ Set-up of geometry}

Let us begin with geometric setup. In \cite{Vafa}, non-supersymmetric vacua in Type IIB string theory were engineered by wrapping D5 branes and anti-D5 branes on $2$-cycles of local Calabi-Yau manifolds. One of the geometries that we will use in this section is 
\begin{equation}
uv=y^2 +W^{\prime}(x)^2.\label{ourgeometry}
\end{equation}
For the sake of simplicity, we study the geometry with two critical points \cite{geometrictransition},
\begin{equation}
W'(x) = g (x-a_1)(x-a_2).
\end{equation}
Here, the coordinates, normalized by the string length $l_s$, are dimensionless. The $2$-cycles at the critical points, we denote $[C_1]$ and $[C_2]$, are not independent but in the same homology class, $[C_1]+[C_2]=0$. So, $N+n$ D5 branes wrapped on $[C_1]$ and $n$ anti-D5 branes wrapped on $[C_2]$ can annihilate with each other, and transit to a supersymmetric (SUSY) vacuum. The SUSY vacua are degenerated and, in fact, there are $N+1$ types of distribution of the remaining D5 branes: $N_1$ D5 branes wrapping on $[C_1]$ and $N-N_1$ on $[C_2]$ where $N_1$ runs from $0$ to $N$. The manifold is engineered so that the potential barrier between the critical points is high enough to avoid the annihilation of D5/anti-D5 branes. The area of $2$-cycle $A(x)$ between them is described by a function of $W^{\prime}(x)$ and $r$,  \cite{Vafa}. 
\begin{equation}
A(x)=l_s^2\sqrt{|W^{\prime}(x)|^2+|r|^2}, \label{twoArea}
\end{equation}
where $r$ is the physical minimal size of the $2$-cycles at the critical points originating from the $B$ field flux on the $2$-cycles, $r=\int_{C_i} B$. Here, we assume $r$ is dimensionless by appropriately normalizing with $l_s$, and we recovered overall $l_s^2$ in \eqref{twoArea} for the later reference. 
Hence, to annihilate with anti-D5 branes, the D5 branes have to first enlarge the size of the wrapped area, which increases the total energy and guarantees the stability of the vacuum. The metastability of the system was discussed from two viewpoints \cite{Vafa}. One is open string description which is reliable when the numbers of wrapped branes are small, and the other is holographic dual description which is reliable when the numbers of branes are large. In \cite{Vafa}, it was conjectured that holographic dual description of the non-supersymmetric configuration can be described by the geometric transition \cite{geometrictransition}. Essentially, the analysis of the decay processes of two cases are similar, so we would like to focus on the open string description below. Since we discuss the non-compact manifold, we have a freedom to shift the constant of the potential like the usual SUSY field theory in four dimension. By choosing $V=0$ for the SUSY vacuum, the vacuum energy of a metastable state is given by 
\begin{equation}
\Delta V=2 {|r|\over g_s l_s^4 }n,
\end{equation}
where $n$ is the number of anti-brane wrapped on $[C_2]$.

In the paper \cite{Vafa}, the decay process of the metastable vacuum was also discussed. In the present assumption, the potential barrier created by the geometric setup is high enough. Hence, one can estimate the process by the thin-wall approximation \cite{Coleman}. From WKB approximation, the decay rate of a false vacuum is estimated by the bounce action $B_{O(4)}$,
\begin{equation}
\Gamma \sim \exp(-B_{O(4)}). \label{vacdecay}
\end{equation}
The bounce action for a bubble with the radius $R$ is
\begin{equation}
B_{O(4)}=-{\pi^2 \over 2} R^4 \Delta V +2\pi^2 R^3 T_{\rm DW},\label{Baction}
\end{equation}
where $T_{\rm DW}$ is the tension of domain wall interpolating between a false vacuum and a true vacuum. The first term comes from the contribution of true vacuum inside the bubble. The second one is energy increase by the domain wall tension. In the current setup, the domain wall is a D5 brane wrapping on the $3$-chain $[B]$ which is the set of two-cycles between $x=a_1$ and $a_2$ \cite{Vafa}. The size of the $3$-chain is 
\begin{equation}
V_3= l_s \int_{a_1}^{a_2} dx \,A(x)~.
\end{equation}
Below, we assume $\Delta \equiv (a_2-a_1)\gg |r|$. So the tension of the domain wall is given by
\begin{equation}
T_{\rm DW}=T_{D5}V_3~.
\end{equation}
By minimizing the action, we obtain the critical size of the bubble, 
\begin{equation}
R_{\rm crit}= 3 {T_{\rm DW} \over \Delta V}.
\end{equation}
Bubbles formed with larger than this size expands without bound and induce the quantum tunneling. Plugging into \eqref{Baction}, the bounce action to estimate the decay rate is given by
\begin{equation}
B_{O(4)}= {27\pi^2  \over 2}  {T_{\rm DW}^4 \over (\Delta V)^3}.\label{bouncemono}
\end{equation}
As long as the $T_{\rm DW}$ is large or $\Delta V$ is small, the vacuum is long-lived.

\section{Fuzzy monopole \label{EnergyMonopole}}

Now we are ready to study monopole-like brane configuration. Consider a D3 brane wrapping on the $3$-chain $[B]$ and ending on D5 branes at $x_1$ and anti-D5 branes at $x_2$. This D3 brane can be seen as a monopole from four dimensional viewpoint\footnote{The closely related metastable stringy monopole in the conifold was initially studied in \cite{StringyMonopole}.}. Without specifying the mechanism, we simply assume the existence of the monopole and study dynamics induced by the monopole. As mentioned in the previous section, the domain walls connecting SUSY vacua and SUSY breaking vacua are D5 branes wrapping also on the $3$-chain $[B_3]$. So the monopole brane is on the top of the domain wall D5 brane. Hence, the D3 branes should dissolve into the domain wall D5 brane and forms energetically favorable bound state \cite{Text}. The energy of the bound state is given by
\begin{equation}
E_{D5/D3}=\sqrt{  \left(T_{D5} 4\pi R^2 V_3 \right)^2+\left(n_{D3} T_{D3} V_3\right)^2 }.
\end{equation}
Here, we assumed that the D5 brane forms spherical shape in $\mathbb{R}^{1,3}$. Since the inside of the wall is filled by the true vacuum, by subtracting the energy gain from the true vacuum, the total energy becomes 
\begin{equation}
E_{\rm total}= \sqrt{  \left(T_{D5} 4\pi R^2 V_3 \right)^2+\left(n_{D3} T_{D3} V_3\right)^2 }-{4\pi \over 3} \Delta V R^3+{\rm const} \label{TotalE},
\end{equation}
where constant term comes from the total vacuum energy of space-time filled by metastable vacuum. Hereafter, we neglect the constant term because it does not play any role for the decay process. As an illustration, we show the radius dependence of the total energy in the figure 1. 
\begin{figure}[htbp]
\begin{center}
 \includegraphics[width=.5\linewidth]{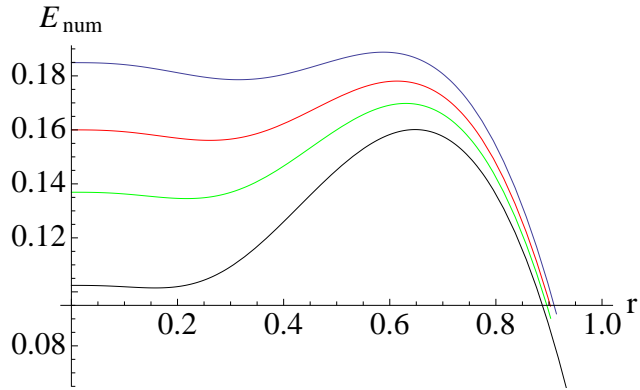}
\vspace{0cm}
\caption{\sl Plots of the dimensionless energy function defined in \eqref{Enum}.  The black, green, red and blue lines correspond to $b=0.32,0.37,0.4$ and $0.43$, respectively.}
\end{center}
\end{figure}
It is useful to introduce the following dimensionless parameters, 
\begin{equation}
E_{\rm total}={36\pi T_{DW}^3\over \Delta V^2}\left[\sqrt{r^4+b^4}-r^3 \right]\equiv    {36\pi T_{DW}^3\over \Delta V^2} E_{\rm num}\label{Enum},
\end{equation}
where
\begin{equation}
r={ \Delta V\over 3T_{DW}}R,\qquad b^4=\left( {\Delta V\over 3T_{DW}}\right)^4 \left({T_{D3}n_{D3}\over 4\pi T_{D5}} \right)^2.
\end{equation}
It is interesting that this energy function is closely similar to the one discussed in \cite{Hashimoto} on fuzzy brane puffed-up by the background flux. Although the physical meaning in the present situation is different, from the similarity, we can proceed calculations of the decay rate in the next section along the lines of \cite{Hashimoto} or \cite{PTK1}. Clearly, for large $b$, the dielectric brane tends to be unstable. The critical value of the instability is  
\begin{equation}
b>{\sqrt{2} \over 3}\qquad {\rm (unstable)}\label{stability}.
\end{equation}
It is remarkable that even if the metastable vacuum (without monopole) is long-lived, the existence of the monopole makes the life-time much shorter: For the parameter choice \eqref{stability}, the vacuum is unstable against inhomogeneous semi-classical decay triggered by expansion of a monopole/bubble bound state. Eventually the original false vacuum is filled by true vacuum in the bubble. 

When $b$ is smaller than the critical value, there is minimum at 
\begin{equation}
r_{\rm min}={1\over 3}\sqrt{2-\sqrt{4-81 b^4}} \label{Rmin}.
\end{equation}
We find that the bound state is energetically favorable, hence domain wall bubble naturally induced by the effect of monopole. This may be regarded as the Schwinger effect induced by the monopole. It is interesting that the stability of the bound state against collapsing is guaranteed by the true vacuum inside the wall, rather than the background flux discussed in \cite{Hashimoto,PTK1,PTK2,hyakutake}. In this sense, our dielectric brane offers a new way to construct a stable fuzzy brane.

\begin{figure}[htbp]
\begin{center}
 \includegraphics[width=.4\linewidth]{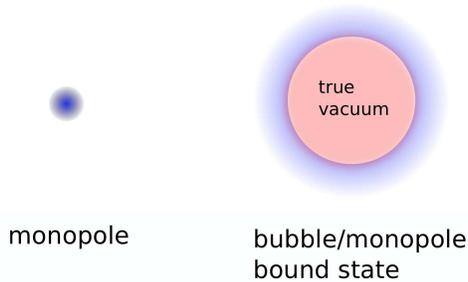}
\vspace{0cm}
\caption{\sl Spontaneous nucleation of bubble triggered by the monopole. The bound state of bubble with the monopole is energetically favorable. When $b>{\sqrt{2}/3}$, the radius of the bubble expands without bound, and the space-time $\mathbb{R}^{1,3}$ is eventually filled by the true vacuum. }
\end{center}
\end{figure}

Finally let us comment on the topological charge carried by the bound state. The original monopole is a D3 brane so there is a coupling to the RR-flux $\int_{D3} C_4$. On the D5 brane there is non-trivial magnetic field $F$ corresponding to the dissolving D3 branes. So there is the Chern-Simon interaction 
\begin{equation}
\int_{D5} C_4\wedge F,
\end{equation}
representing that the bound state carries the original charge of $C_4$. In this sense, D5/D3 bound state is a bubble with the monopole winding number\footnote{In a broad sense, this may be regarded as an example of soliton bubbles discussed in \cite{John}.}.

\section{Decay rate of metastable monopole \label{DecayMonopole}}

\begin{figure}[htbp]
\begin{center}
 \includegraphics[width=.5\linewidth]{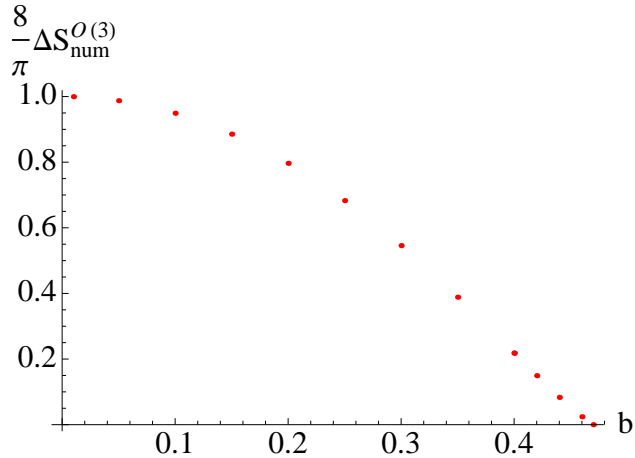}
\vspace{-.1cm}
\caption{\sl Numerical estimation of dimensionless bounce action defined in \eqref{dimlessbounce}.}
\label{FIGnumBmono}
\end{center}
\end{figure}

Now, let us tune the parameter $b$ such that dielectric branes are metastable. In this case, semi-classical rolling over does not take place, instead, $O(3)$ symmetric bubble is created and quantum tunneling occurs. To estimate the decay rate, we follow Coleman's method \cite{Coleman}. The effective action of D-brane is given by the Dirac-Born-Infeld (DBI) action, so we will study the bounce solutions of Euclidean DBI action for estimation of decay rate. Basically, we can proceed along the lines of \cite{Hashimoto,hyakutake,PTK1,PTK2} where decay process of unstable dielectric brane created by background flux were studied. We consider D5 brane corresponding to the domain wall connecting metastable vacuum and true vacuum. D3 monopoles are dissolved, so, magnetic flux is generated on the D5 brane. Suppose that embedding function of the D5 brane is spherically symmetric,
\begin{eqnarray}
X^0=t,\quad X^1=R(t) \sin \theta \cos \varphi ,\quad X^2=R(t) \sin \theta \sin\varphi ,\quad X^3=R(t) \cos \theta ,
\end{eqnarray}
in the $ \mathbb{R}^{1,3}$. Also, $X^{4,5,6}$ are filling the three-chain $[B]$. The others are constant. The magnetic flux on the brane exists along $(\theta , \varphi)$ directions, $B\sin \theta \equiv 2\pi \alpha^{\prime}F_{\theta \varphi} $, hence we obtain
\begin{eqnarray}
-\det (\partial_{\alpha}X^{\mu}\partial_{\beta} X_{\mu}+2\pi \alpha^{\prime}F_{\alpha \beta})&=& -\det (\partial_{a}X^{\mu}\partial_{b}X_{\mu} +2\pi \alpha^{\prime}F_{\alpha \beta})\cdot \det (\partial_{A}X^{I}\partial_{B} X_{I} ) \nonumber \\
&=&\sin^2\theta(1-\dot{R}^2)(R^4+B^2)  \cdot \det (\partial_{A}X^{I}\partial_{B} X_{I} ) ,\label{determinant}
\end{eqnarray}
where $I=4,5,6$ and $a,b=(t,\theta ,\varphi)$. $A,B$ run for the coordinates of $[B]$. Since $X^{4,5,6}$ depend only on the coordinates of $[B]$ and $X^{0,1,2,3}$ depend only on those of $ \mathbb{R}^{1,3}$, there is no cross term in the matrix of the left-hand side in \eqref{determinant}. Hence the matrix is block diagonal form and the determinant of the matrix becomes the product of two pieces. With this relation, the DBI action is given by
\begin{eqnarray}
S&=& T_{D5}\int d^6\xi  \sqrt{\det (\partial_{A}X^{\mu}\partial_{B} X_{\mu} )}\sqrt{-\det (\partial_{\alpha}X^{\mu}\partial_{\beta} X_{\mu}+2\pi \alpha^{\prime}F_{\alpha \beta})}- \int dt\  {4\pi \over 3}R^3 \Delta V \nonumber \\
&=& \int d t\left[  4\pi T_{D5}V_3  \sqrt{(1-\dot{R}^2 ) (R^4 +B^2)  } -   {4\pi \over 3}R^3 \Delta V\right], \label{DBID5}
\end{eqnarray}
where we defined
\begin{equation}
V_3=\int d^3 \xi^A  \sqrt{\det (\partial_{A}X^{I}\partial_{B} X_{I} )}.
\end{equation}
This is the volume of the $3$-chain $[B]$ in the approximation $r\ll 1$. Comparing \eqref{DBID5} with \eqref{TotalE}, we can easily find the following relation, $B=n_{D3} T_{D3}/ 4\pi  T_{D5}$. It is convenient to introduce dimensionless parameters for the later numerical computation of the decay rate. 
\begin{equation}
r={ \Delta V\over 3T_{DW}}R,\qquad b^4=\left( {\Delta V\over 3T_{DW}}\right)^4 B^2, \qquad s={ \Delta V\over 3T_{DW}}\tau.
\end{equation}
By Wick rotation, we introduce the Euclidean time $t\to i\tau $. The Euclidean action is written as follows:
\begin{eqnarray}
S_E&=&\left( {27\pi^2\over 2} {T_{\rm DW}^4 \over \Delta V^3} \right) {8\over \pi} \left[\int ds \left(- \sqrt{(r^4+b^4)(1+\dot{r}^2)}  +r^3 \right) \right] \nonumber \\
&\equiv & B_{O(4)} {8\over \pi} S_{\rm num}^{O(3)}. \label{dimlessbounce}
\end{eqnarray}
Here, we used the bounce action \eqref{bouncemono} for $O(4)$ symmetric bubble. Below, we study bounce solutions of this action and estimate the decay rate numerically. The exponent of the decay rate is the difference between the Euclidean action of the bounce solution and the action of the background 
\begin{equation}
B_{O(3)}= B_{O(4)} {8\over \pi}\left[S^{O(3)}_{\rm num}(r_{\rm bounce})-S^{O(3)}_{\rm num}(r_{\rm min}) \right] =B_{O(4)}{8\over \pi} \Delta S_{\rm num}^{O(3)},
\end{equation}
where $r_{\rm bounce}$ represents the bounce solution. We find that exact analysis of $\Delta S_{\rm num}^{O(3)}$ is involved. So we will show the numerical estimation, which is enough for our purpose. We are looking for the solution starting from $r_{\rm min}$ and goes until $r_{\rm max}$ and coming back to the original position. The equation of motion can be represented by the first order differential equation, 
\begin{equation}
\partial_s \left[  \sqrt{r^4+b^4 \over 1+\dot{r}^2}  -r^3 \right]=0. 
\end{equation}
By the integration constant $C$, one can write the solution of the equation easily, and find that the velocity of the bounce is given by, 
\begin{equation}
 {dr\over d s} = \pm {1\over C+r^3}\sqrt{ r^4+b^4-(C+r^3)^2 } \label{velocity}.
\end{equation}
Until $r_{\rm max}$, the velocity is positive, so the solution must be satisfied the following condition, 
\begin{equation}
K(b)\equiv r^4+b^4-(C+r^3)^2=(r-r_{\rm min})(r_{\rm max}-r)(r^4+a_3r^3 +a_2r^2 +a_1 r+a_0) \label{constraint}.
\end{equation}
The minimum value is already obtained in \eqref{Rmin}, hence, six equations arising from \eqref{constraint}, give us six variable as functions of $b$,  $a_i(b)$, $C(b)$, $r_{\rm max}(b)$. Here, we solve this condition numerically and carry out the following integration. By using \eqref{velocity}, the integration for $s$ can be expressed as that of $r$, 
\begin{eqnarray}
\Delta S_{\rm num}^{O(3)}&=&\int_{r_{\rm min}}^{r_{\rm max}} dr \left[ {C+r^3\over \sqrt{K(b)}}\left( {r^4+b^4 \over C+r^3}-r^3 \right)  -{C+r^3 \over \sqrt{K(b)}}\left(\sqrt{r^4_{\rm min} +b^4 } -r_{\rm min}^3\right) \right]  \nonumber  \\
&= &\int_{r_{\rm min}}^{r_{\rm max}} dr \sqrt{K(b)},
\end{eqnarray}
where we used the integration constant evaluated at $s_{\rm initial}$,
\begin{equation}
C=\sqrt{r_{\rm min}^4+b^4} -r_{\rm min}^3.
\end{equation}
Numerical result of this integral with $8/\pi $ factor is presented in the figure \ref{FIGnumBmono}. As expected, in the limit $b\to 0$ (no monopole limit), the decay rate goes back to $O(4)$ symmetric result. Under the zero magnetic flux, $O(3)$ symmetric bounce is not the most effective bounce action, $O(4)$ symmetric one is the most effective. This gives a test for the argument \cite{Coleman2} for the case of DBI action. In the DBI action, the kinetic term and potential term is not separated, so our result is non-trivial check of the argument. Also, the results is consistent with the fact that for $b\ge \sqrt{2}/3$, the monopole becomes unstable and semiclassical decay occurs rather than quantum tunneling.

\section{Topology change of bubble \label{DecayDyon}}

So far, we have studied the bound state of the spherical bubble and the monopole. Here we will present that colliding fundamental strings to the bound state induces a topology change of the spherical bubble. As pointed out in \cite{NishiokaTakayanagi} in the context of the giant graviton, colliding fundamental strings can dissolve into a brane and generate the electric flux on it. Then, non-zero pointing vector by electric/magnetic flux induces angular momentum near the dissolving string, and eventually a spherical giant graviton turns to a tube-like one. We apply their idea to our soliton bubble. As we will see below, the spherical anzats of monopole-like bubble with electric flux is not appropriate. So we expect that the topology of the bubble changes to the torus. 

First of all, let us naively consider the spherical anzats of monopole/bubble bound state with electric flux. Turning on the electric field on the sphere $E_{\theta}=2\pi \alpha^{\prime}F_{t,\theta }$, we obaint an extra contribution in the DBI action
\begin{equation}
S= 4\pi T_{D5}V_3 \int d t \sqrt{(R^4+B^2)(1-\dot{R}^2)+E^2R^2 } - \int dt\  {4\pi \over 3}R^3 \Delta V.
\end{equation}
By Legendre transformation, we can rewrite the action by using the electric density flux $D\equiv {\partial L \over \partial E}$,
\begin{equation}
D={4\pi T_{D5}V_3  ER^2\over \sqrt{(R^4+B^2)(1-\dot{R}^2)+E^2R^2 }}\equiv 4\pi T_{D5}V_3 \widetilde{D},
\end{equation}
\begin{equation}
{ S}=\int d t \left[ 4\pi T_{DW}   {1 \over
	       R}\sqrt{(R^4+B^2)(1-\dot{R}^2)(R^2-\widetilde{D}^2)} -
	       {4\pi \over 3}R^3 \Delta V\right].\\
\end{equation}
Clearly, one find that this action is not well-defined for small radius $R$. So we assume that a topology change of bubble has to happen, and torus-like tube, which is consistent with the existence of electric flux (equivalently angular momentum), is expected to be formed after the transition. As is schematically shown in the figure 4, the fundamental string ending on the monopole creates the handle to the bubble and by the angular momentum generated at the attached points blows up the size of the handle, and eventually stable torus would be formed. To investigate this transition precisely, we have to study the time dependent solution which is quite difficult to solve. Thus, below, we simply assume the torus with length $L$ is formed after the transition and estimate the tunneling rate of the configuration numerically.

\begin{figure}[htbp]
\begin{center}
 \includegraphics[width=.5\linewidth]{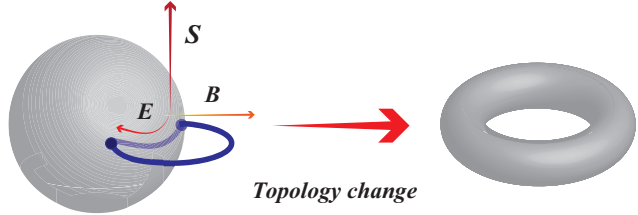}
\vspace{0cm}
\caption{\sl Topology change of bubble}
\end{center}
\end{figure}

\begin{figure}[htbp]
\begin{center}
 \includegraphics[width=.5\linewidth]{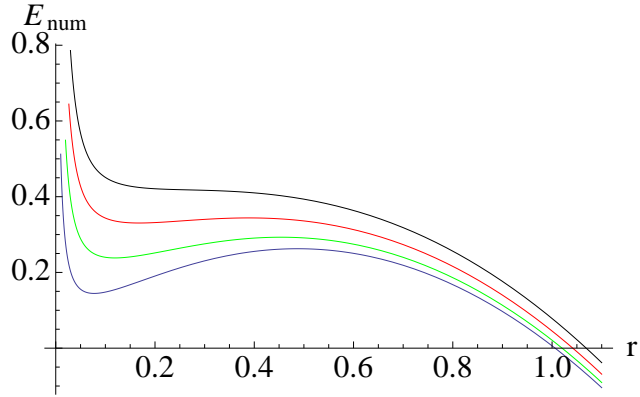}
\vspace{-.1cm}
\caption{\sl $b=1/20$, $d=1/10,2/10,3/10,4/10$ for blue, green, red and black.}
\label{FIGpotdyon}
\end{center}
\end{figure}

As the embedding function of the tube-like D5 brane, we assume 
\begin{eqnarray}
&&X^0=t, \qquad X^1=z, \qquad X^2=R(t)\cos \theta ,\qquad X^3=R(t)\sin \theta  \nonumber ,
\end{eqnarray}
the others are constant. The electric flux is along the $z$ direction, and magnetic flux is $(\theta, z)$ directions, so the total action for the D5 brane becomes, 
\begin{equation}
S=\int d t\left[ 2\pi L T_{D5} V_3 \sqrt{R^2(1-\dot{R}^2-E_z^2)+B^2(1-\dot{R}^2)} - \pi  R^2 L \Delta V\right].
\end{equation}
For the sake of simplicity, we assume that the radius of the tube is
the same along the $z$-direction. Also, the length $L$ is much larger
than the radius of the tube to neglect the effects of the curvature of
the tube. Again, take the Wick rotation of DBI action. By introducing
the electric density flux, ${\delta {\cal L}_{E}\over \delta
E_{z}^{\rm euc}}\equiv 2\pi LT_{\rm DW}\widetilde{D}_{E}^{\rm euc}$,
the action is given by
\begin{equation}
 S_{ E}=\int d\tau \left[ 2\pi L T_{D5} V_3  {-1\over R}\sqrt{(R^2+B^2)(R^2+(\widetilde{D}_E^{\rm euc})^2)(1+\dot{R}^2)}+\pi  R^2 L \Delta V\right] \nonumber \\
\end{equation}
 \begin{eqnarray}
&=& 8\pi L {T_{\rm DW}^3 \over \Delta V^2} \int ds \left[ {-1\over r}\sqrt{(r^2+b^2)(r^2+d^2)(1+\dot{r}^2)}  +r^2  \right]\nonumber \\
&\equiv &8\pi L {T_{\rm DW}^3 \over \Delta V^2} S_{\rm num}^{O(2)},
 \end{eqnarray}
where 
\begin{equation}
r= { \Delta V\over 2T_{\rm DW}} R,\quad s= { \Delta V\over 2T_{\rm DW}}\tau ,\quad b=B { \Delta V\over 2T_{\rm DW}}, \quad d=\widetilde{D}_E^{\rm euc} { \Delta V\over 2T_{\rm DW}}.
\end{equation}
For the static case $\dot{R}=0$, as shown in the figure \ref{FIGpotdyon},  there is minimum away from the origin. This is because the nonzero angular momentum blows up the tube rather than true vacuum insider the bubble \cite{Supertube}. In fact, there still exists non-zero minimum in the limit $\Delta V\to 0$.

\begin{figure}[htbp]
\begin{center}
 \includegraphics[width=.5\linewidth]{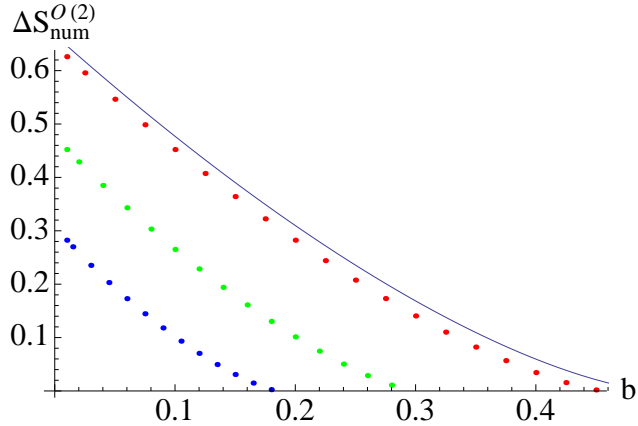}
\vspace{-.1cm}
\caption{\sl Numerical estimation of dimensionless bounce action defined in \eqref{dimlessbounce2}. Blue solid line corresponds to $d=0$. The red, green and blue dots correspond to $d=1/100,1/10$ and $2/10$ respectively.}
\label{FIGnumBdyon}
\end{center}
\end{figure}

Let us estimate the decay rate in this case. From the Euclidean action, one obtain the first differential equation,
\begin{equation}
\partial_s \left(   {1\over r} \sqrt{(r^2+b^2)(r^2+d^2)\over (1+\dot{r}^2)}  -r^2 \right)=0.
\end{equation}
This equation can be easily solved 
\begin{equation}
\dot{r}=\pm {\sqrt{(r^2+b^2)(r^2+d^2)-r^2(c+r^2)^2}\over r(c+r^2)}.
\end{equation}
The bounce solution has positive velocity between $r=r_{\rm min}$ and $r_{\rm  max}$, the following factorization condition should be satisfied,
\begin{equation}
(r^2+b^2)(r^2+d^2)-r^2(c+r^2)^2=(r^2-r_{\rm min}^2)(r_{\rm max}^2-r^2)(r^2+a_0).
\end{equation}
The exponent of the tunneling rate is given by the subtracted bounce action with the original action 
\begin{eqnarray}
B_{O(2)}&=&16\pi L {T_{\rm DW}^3 \over \Delta V^2} \int_{r_{\rm min}}^{r_{\rm max}} {r(c+r^2)dr\over \sqrt{(r^2-r_{\rm min}^2)(r_{\rm max}^2-r^2)(r+a_0)}}\times \nonumber \\
&&\qquad \left[ {(r^2+b^2)(r^2+d^2)\over r^2(c+r^2)}-r^2-\left( {\sqrt{(r_{\rm min}^2+b^2)(r_{\rm min}^2+d^2)}\over r_{\rm min}}-r_{\rm min}^2 \right) \right].
\end{eqnarray}
The factor $2$ comes from the round trip of the solution. The integration constant $c$ can be written by using $r_{\rm min}$
\begin{equation}
c={\sqrt{(r_{\rm min}^2+b^2)(r_{\rm min}^2+d^2)}  \over r_{\rm min}}  -r_{\rm min}^2 .
\end{equation}
With this relation, and introducing new variable, $x=r^2$, the bounce action is represented as 
\begin{eqnarray}
B_{O(2)}&=&16\pi L {T_{\rm DW}^3 \over \Delta V^2} \int_{x_{\rm min}}^{x_{\rm max}}{dx \over 2x}\sqrt{(x-x_{\rm min})(x_{\rm max}-x)(x+a_0)} \nonumber \\
&=& 8\pi L {T_{\rm DW}^3 \over \Delta V^2} \Delta S_{\rm num}^{O(2)}. \label{dimlessbounce2}
\end{eqnarray}
Numerical estimations of this integral for various parameters are shown in figure \ref{FIGnumBdyon}. As for the case with $d=0$, we find that the solution is exactly obtained
\begin{eqnarray}
B_{O(2)}(d=0)&=&16\pi L {T_{\rm DW}^3 \over \Delta V^2} \int_0^{r_{\rm max}^0} 2r  \sqrt{(r_{\rm max}^0)^2-r^2}  dr \nonumber \\
&=&8\pi L {T_{\rm DW}^3 \over \Delta V^2}\cdot {2\over 3} \left(1-2b\right)^{3\over 2},
\end{eqnarray}
where $r_{\rm  max}^0=\sqrt{1-2b}$. 

After the topology change, the length of the torus should be at least larger than the size of original monopole. The minimum of the monopole is given by
\begin{equation}
R_{\rm mono} ={3T_{\rm DW} \over \Delta V}r_{\rm min}.
\end{equation}
By normalizing with this scale, the length of the torus is given by
\begin{equation}
l\equiv {L\over R_{\rm mono}}.
\end{equation}
With the dimensionless parameter, the bounce action for $O(2)$ symmetric bubble is given by
\begin{eqnarray}
B_{O(2)} =B_{O(3)} {2r_{\rm min} \Delta S_{\rm num}^{O(2)}\over 9 \Delta S_{\rm num}^{O(3)}}l.
\end{eqnarray}
When the condition,
\begin{equation}
l > { 9 \Delta S_{\rm num}^{O(3)}\over 2r_{\rm min} \Delta S_{\rm num}^{O(2)}},
\end{equation}
is satisfied, the decay rate becomes smaller than that of $O(3)$ bubble. Thus the topology change the bubble makes the life-time of the bubble longer. Also, it is interesting that shorter loops than this critical length have naively larger bounce action which means $O(2)$ symmetric decay is faster than that of $O(3)$ case. However, in this region, our analysis loose control and there should exist significant corrections originating from curvature effect of the torus. So, more careful treatment is required.  

\section{Warped compactification}

Finally, in this section, we comment on effects of the warp factor. Let us consider the geometry in the previous sections with the warp factor $e^{2A}\ll 1$,
\begin{equation}
ds^2 =e^{2A(y)} g_{\mu \nu}dx^{\mu}dx^{\nu}+e^{-2A(y)} g_{mn}dy^m dy^n.
\end{equation}
$g_{mn}$ represents the metric of the geometry used in earlier discussions. To the best of our knowledge, the explicit metric corresponding to the generalized geometry \eqref{ourgeometry} has not been known yet. So here we show naive arguments by focusing on the case where $A(y)$ is almost constant. Accounting for the factor, the volume of the three-chain (we denote $V^{\rm warp}$) on which the monopole and the domain walls are wrapping is given by 
\begin{equation}
V^{\rm warp}=V (e^{-2A})^{3/2},
\end{equation}
where $V$ is the volume of the three-chain for the non-warped geometry. Hence, the tension of the domain wall is also modified as  
\begin{equation}
T^{\rm warp}_{\rm DW}=T_{D5}V^{\rm warp}e^{3A}.
\end{equation}
Following the same argument shown in section 2, we find that the
tension of the wall doesn't change with warping but the energy drop becomes smaller. So totally, the bounce action becomes much larger than that of non-warped geometry
\begin{equation}
B_{O(4)}^{\rm warp}=B_{O(4)} e^{-6A}.
\end{equation}
Therefore, the warping effect makes the life-time of the vacuum longer. 

However, remarkably as for the decay process via $O(3)$ symmetric bubble induced by monopoles, conclusion is opposite. The relevant term in DBI action is
\begin{equation}
T_{\rm DW}^{\rm warp}\int d^3 \xi   \sqrt{-\det (\partial_{\alpha}X^{\mu}\partial_{\beta}X_{\mu} +e^{-4A}B^2)}.
\end{equation}
To estimate the ratio of $B_{O(3)}^{\rm warp}$ and $B_{O(4)}^{\rm warp}$, one can use the formulae presented in the section 4 by the following replacement, 
\begin{equation}
T_{\rm DW}\to T_{\rm DW}^{\rm warp},\qquad V \to V^{\rm warp}, \qquad B^2\to B^2e^{-4A}.
\end{equation}
From this, we immediately conclude that when non-zero magnetic field is turned on, by the warp factor the effective strength of magnetic field is largely enhanced and the monopole/bubble bound state becomes unstable, 
\begin{equation}
e^{-2A}b> {\sqrt{2}\over 3} \qquad {\rm (unstable)}.
\end{equation}
In a wide range of parameter choices, this instability condition is easily satisfied. Thus, in the warped compactification, the vacuum decay via $O(3)$ symmetric bubble induced by the monopole is significantly important.

\section{Conclusions and discussions}

In this paper, we have studied dynamics of dielectric branes in false vacua in Type IIB string theory. We have seen that the dielectric branes were either unstable or metastable depending on the strength of magnetic flux induced by the dissolving D3 brane (corresponding to a monopole). As for the metastable monopole configuration, the lower energy vacuum filling in the spherical bubble generates a force enlarging the size of the bubble and creates finite size of a bubble/monopole bound state. This is a new mechanism for stabilizing the dielectric branes without using background flux nor angular momentum. We showed that existence of the solitonic branes makes creation probability of $O(3)$ or $O(2)$ symmetric bubbles large, and makes the life-time of the false vacuum much shorter. As for the unstable monopole, the decay process was semi-classical rather than quantum mechanical since it does note require the quantum tunneling. Also, we showed that topology change of the soliton bubble triggered by colliding fundamental strings, which makes the tunneling rate small in a wide range of parameter space. We claim that these remarkable phenomena significantly affect the life-time of false vacua and the evolution of created bubbles, hence affect scenarios in the early universe based on the string landscape. 

\begin{figure}[htbp]
\begin{center}
 \includegraphics[width=.5\linewidth]{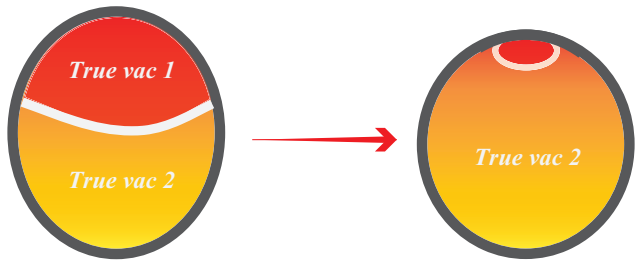}
\vspace{0cm}
\caption{\sl Two SUSY preserving vacua filling in the dielectric brane. The energy can be reduced by making one of the regions small.}
\label{FIGdomainwall}
\end{center}
\end{figure}

It would be fascinating if we could extract information of other vacua from the outside of the bubble. In our concrete set-up, there are degenerated supersymmetry preserving vacua. Hence, there is a possibility that a part of a dielectric brane is filled by one SUSY vacuum and the other part by another SUSY vacuum. In this case, between the two regions, there should be a domain wall connecting two different vacua. For example, for the monopole-like dielectric brane, the configuration would be no longer spherical but slightly modified due to the domain wall, see figure \ref{FIGdomainwall}. However, since this configuration is not stable, one of the regions may shrink and eventually disappear (in the right of figure 7). So, it may be hard, though it is not impossible in principle, to get information from outside of the bubble.

Finally, we would like to comment on gravitational effects. In this paper, for the sake of simplicity, we focussed on a compactification on a non-compact manifold and took the small $g_s$ limit. Hence, both four-dimensional and ten-dimensional gravitational effects were neglected at the leading order. However, in realistic model building, we have to deal with a compact manifold and include the effects of gravity. In studying such situations, we first have to take care of moduli fixing corresponding to the parameters in the present non-compact geometry. Also, for bubble formation and expansion, we have to reconsider the bounce solution for the DBI action along the lines of arguments by Coleman and De Luccia \cite{CD}. These issues are interesting but beyond the scope of this paper, so we would like to leave them for future work. Also, to show the generosity of our idea, it would be important to extend current study further to similar but different contexts such as a cosmic string \cite{EHT} in metastable configurations in Type IIA string theory \cite{OoguriOokouchi} or solitons in metastable state in Type IIB theory compactified on $A_n$ geometries \cite{A2} (see \cite{KOOreview} for a recent review). Furthermore, studying inhomogeneous decay of metastable vacua in perturbed Seiberg-Witten theory \cite{SeibergWitten} would be interesting since there is a massive `t Hooft-Polyakov monopole in this theory which can be a trigger for the vacuum decay. We would like to discuss these topics in separate publications.

\section*{Acknowledgement}

The authors are grateful to Yoshifumi Hyakutake, Tatsuo Kobayashi and Yoske Sumitomo for useful comments and discussions. YO would like to thank Shigeki Sugimoto for discussions on the monopole in the geoemetry used in this paper. The authors would like to thank Hokkaido University and the organizers of ``KEK Theory Workshop 2015'' for their kind hospitality where this work was at the final stage. This work is supported by Grant-in-Aid for Scientific Research from the Ministry of Education, Culture, Sports, Science and Technology, Japan (No. 25800144 and No. 25105011).

%
%

\end{document}